\documentclass[fleqn,usenatbib]{mnras}

\usepackage{newtxtext,newtxmath}

\usepackage[T1]{fontenc}

\DeclareRobustCommand{\VAN}[3]{#2}
\let\VANthebibliography\thebibliography
\def\thebibliography{\DeclareRobustCommand{\VAN}[3]{##3}\VANthebibliography}

\usepackage{graphicx}	
\usepackage{amsmath}	
\usepackage{subfig}

\newcommand{\hMpc}{{\ifmmode{~h^{-1}{\rm Mpc}}\else{$h^{-1}$Mpc}\fi}}
\newcommand{\R}{{\ifmmode{~$R_{200}$}\else{$R_{200}$}\fi}}
\newcommand{\Msun}{{\ifmmode{{\rm {M_{\odot}}}}\else{${\rm{M_{\odot}}}$}\fi}}
\newcommand{\hMsun}{{\ifmmode{~h^{-1}{\rm M_{\odot}}}\else{$h^{-1}{\rm{M_{\odot}}}$}\fi}}
\newcommand{\gadgetx}{\textsc{Gadget-X}}
\newcommand{\threehundred}{\textsc{The ThreeHundred }}
\newcommand{\disperse}{\textsc{DisPerSE}}

\defcitealias{Kuchner2020}{Paper~I}

\title[Finding filaments in redshift space]{Cosmic filaments in galaxy cluster outskirts: quantifying finding filaments in redshift space}
\author[U. Kuchner et al.]{\parbox{\textwidth}{
Ulrike Kuchner,$^{1}$\thanks{E-mail: ulrike.kuchner@nottingham.ac.uk (UK)}
Alfonso Arag\'{o}n-Salamanca,$^{1}$
Agust\'{i}n Rost,$^{2}$
Frazer R. Pearce,$^{1}$
Meghan E. Gray,$^{1}$
Weiguang Cui,$^{3}$
Alexander Knebe, $^{4,5,6}$
Elena Rasia,$^{7,8}$
Gustavo Yepes$^{4,5}$
}
\vspace{0.4cm}
\\
\parbox{\textwidth}{
$^{1}$School of Physics \& Astronomy, University of Nottingham, Nottingham NG7 2RD, UK\\
$^{2}$Instituto de Astronomía Te\'orica y Experimental (IATE), Laprida 854, C\'ordoba, Argentina\\
$^{3}$Institute for Astronomy, University of Edinburgh, Royal Observatory, Edinburgh EH9 3HJ, United Kingdom\\
$^{4}$Departamento de F\'isica Te\'{o}rica, M\'{o}dulo 15, Facultad de Ciencias, Universidad Aut\'{o}noma de Madrid, 28049 Madrid, Spain\\
$^{5}$Centro de Investigaci\'{o}n Avanzada en F\'{\i}sica Fundamental (CIAFF), Facultad de Ciencias, Universidad Aut\'{o}noma de Madrid, 28049 Madrid, Spain\\
$^{6}$International Centre for Radio Astronomy Research, The University of Western Australia, 35 Stirling Highway, Crawley, Western Australia 6009, Australia\\
$^{7}$Osservatorio Astronomico di Trieste, Istituto Nazionale di Astrofisica, Via Tiepolo, 11, I-34131 Trieste, Italy\\
$^{8}$Institute for Fundamental, Physics of the Universe, Via Beirut 2, 34014 Trieste, Italy
}}

\date{Accepted 2021 February 22. Received 2021 February 05; in original form 2020 September 02}

\pubyear{2021}

\begin{document}
\label{firstpage}
\pagerange{\pageref{firstpage}--\pageref{lastpage}}
\maketitle

\begin{abstract}
Inferring line-of-sight distances from redshifts in and around galaxy clusters is complicated by peculiar velocities, a phenomenon known as the "Fingers of God" (FoG).
This presents a significant challenge for finding filaments in large observational data sets as these artificial elongations can be wrongly identified as cosmic web filaments by extraction algorithms.
Upcoming targeted wide-field spectroscopic surveys of galaxy clusters and their infall regions such as the WEAVE Wide-Field Cluster Survey motivate our investigation of the impact of FoG on finding filaments connected to clusters. 
Using zoom-in resimulations of 324 massive galaxy clusters and their outskirts from \threehundred project, we test methods typically applied to large-scale spectroscopic data sets. 
This paper describes our investigation of whether a statistical compression of the FoG of cluster centres and galaxy groups can lead to correct filament extractions in the cluster outskirts.
We find that within 5\R\ ($\sim15\hMpc$) statistically correcting for FoG elongations of virialized regions does not achieve reliable filament networks compared to reference filament networks based on true positions.
This is due to the complex flowing motions of galaxies towards filaments in addition to the cluster infall, which overwhelm the signal of the filaments relative to the volume we probe.
While information from spectroscopic redshifts is still important to isolate the cluster regions, and thereby reduce background and foreground interlopers, we expect future spectroscopic surveys of galaxy cluster outskirts to rely on 2D positions of galaxies to extract cosmic filaments.

\end{abstract}

\begin{keywords}

large-scale structure of Universe -- 
galaxies: clusters: general -- 
galaxies: distances and redshifts --
cosmology: observations --
methods: data analysis --
methods: observational
\end{keywords}



\section{Introduction}
\begin{figure*}
   \centering
   \includegraphics[width=\textwidth]{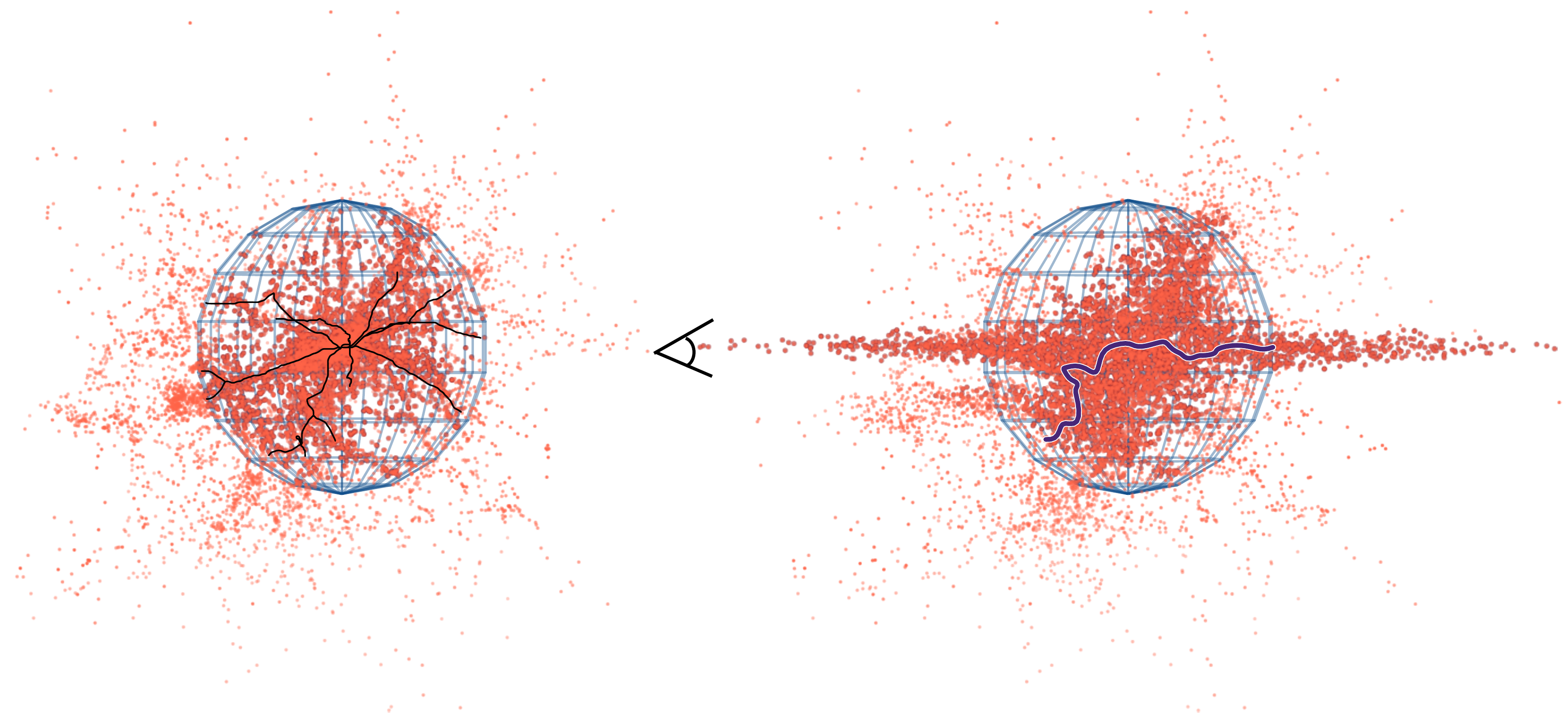}
    \caption{The ideal case in which the filament extraction is based on knowledge of three dimensional positions (left panel) is not accessible through observations. Instead, significant redshift distortions that increase towards the centre of the cluster complicate the extraction of filaments around clusters. Left: true 3D positions of mock galaxies from \threehundred project tailored to WEAVE observations with the reference filament network in black extracted using the \disperse\ software. Right: redshift distorted galaxy distribution with filaments using identical extraction steps for illustration. The black symbol indicates the position of an observer. The blue mesh shows a sphere of 5$R_{200}$ radius, equivalent to roughly $15\hMpc$, centred on the main halo of the cluster. Halos positioned inside this sphere are plotted with larger symbols.}
   \label{fig:problem}
\end{figure*}
The present day Universe is pervaded by a network of filaments that connect groups and clusters brimming with galaxies. In standard cosmology, this highly anisotropic distribution of matter on large scales is the natural consequence of a  hierarchical assembly under the effect of gravity.
The structure develops from the anisotropic gravitational collapse of initial density perturbations \citep{zeldovich1970, Bond1996} building the backbone of the cosmic web. The contrast of the Universe increases with time as rich overdensities grow in mass and density at the intersection of filaments while contracting in physical size. This is the environment in which galaxy clusters form, grow and continue to grow as ultimate manifestations of hierarchical structure formation through a series of mergers and accretion from the cosmic web. Comparatively empty voids expand accordingly, growing to dominate the overall volume in the Universe. 
%
In the Zel'dovich approximation, anisotropic collapse has a well-defined sequence, whereby regions first compress to form walls, then filaments, before finally collapsing along each direction to form clusters \citep{Lin1965,Arnold1982, Shandarin1984, Shandarin1989, Hidding2013, Cautun2012}. 
This general view of structure formation is strengthened by comparing results from numerical simulations that have implemented $\Lambda$CDM cosmological models \citep{Bond1983,Doroshkevich1984} to observations. In a series of successes, ever improving cosmological simulations \citep[e.g.,][]{springel05, Klypin2011, Vogelsberger2014} have been able to describe the formation and evolution of large-scale structures that largely match the observed Universe from galaxy surveys on comparable scales \citep[e.g.,][]{Lapparent1986, Colless2001, Alpaslan2013}. 

Today, the majority of mass relative to the volume occupied in the Universe lies in the small regions of clusters and groups \citep{Tempel2014, Cautun2012, Veena2019}. More specifically, X-ray observations of the hot intracluster medium have located the bulk of gas to just beyond their virial radius and, importantly, within the filaments that connect clusters to the cosmic web \citep[see][for a recent summary]{Walker2019}. Consequently, cosmic filaments are fundamental in transporting both dark matter and baryonic matter into clusters \citep{Cautun2012, Kraljic2018}. The outskirts of galaxy clusters are thus areas of increasing interest for both cosmology and astrophysics. 
Cosmological simulations of galaxy cluster formation depict cluster outskirts as the playing field for large-scale structure formation as it happens: the rich, thermal, kinematic and chemical content of merging sub-clusters, infalling groups, clumps of gas and gas trapped in collapsed dark matter halos funnelled through cold filamentary streams are part of yet-to-be explored accretion physics \citep{ Dekel2009, Danovich2012, Welker2019,Walker2019}.

Mapping cluster outskirts and identifying cosmic filaments connected to clusters, however, is not a trivial task, since the cosmic web comprises a wide range of scales and densities that lead to a plethora of spatial patterns and morphologies. Filaments connected to clusters are only one aspect of a complex multiscale picture that encompasses thick filaments as well as thin tendrils on scales of a few Mpc up to 100 Mpc and more, as well as sheetlike membranes easily mistaken as filaments in projection.
The last decade has seen a number of excellent methods to identify and classify features of the cosmic web \citep[e.g.,][]{Aragon-Calvo2007a, Sousbie2011, Cautun2012, Courtois2013, Tempel2014,Falck2015}, each designed to tackle specific problems to be applied to different kinds of data and thus disagreements are understandable and well documented \citep[e.g.,][]{Libeskind2017, Rost2020}.
Structure finding methods are successfully being applied to simulated and observed datasets alike, including photometric and spectroscopic surveys such as SDSS \citep{Tempel2013, Kuutma2017, chen2016, Malavasi2020}, COSMOS \citep{Darvish2017, Laigle2017} and GAMA \citep{Kraljic2018,Welker2019} amongst others \citep{Malavasi2016,Sarron2019,Santiago-Bautista2020}. The main goal of many of these studies is to investigate galaxy properties such as mass, colour, morphology and gas content, and, increasingly so, alignments and spin orientation with respect to cosmic filaments. 
The interest is based on the well established finding that almost every observable property of galaxies correlates with both galaxy mass \textit{and} galaxy environment, where environment can be defined in various ways, e.g., through local densities, as clusters vs. field, or cosmic web features. Determining and understanding the interplay of the physical processes behind this finding remains a fundamental challenge in understanding galaxy formation and evolution.
The difficulty lies in disentangling subtle competing processes that act on different timescales \citep[e.g., AGN feedback;][]{croton06}, with different mass \citep[e.g., starvation of gas supply;][]{larson80}, and environmental dependence \citep[e.g., ram-pressure stripping;][]{gunngott72}. 
In general, large-scale environmental effects are typically found to be small compared to mass- and local density driven processes and thus need to be controlled carefully for both stellar mass \citep[e.g.,][]{baldry06}, and environment \citep[e.g.,][]{Peng2010} if progress is to be made.
 
In practice, defining and controlling for \textit{environment} is not straightforward due to different definitions, detection methods and treatments, and could be the cause for some perceived disparity of reported results and interpretations \citep[see][for a detailed discussion]{Libeskind2017}.
While modern hydrodynamical cosmological simulations are imperative to provide a census, bridging between simulations and observations requires the full understanding of the performance of observations in realistic setups.
This paper represents a sequel to a paper devoted to the problem of mapping and characterizing filaments around galaxy clusters \citep[][from here on Paper~I]{Kuchner2020}. In \citetalias{Kuchner2020}, we used simulations to discuss strategies and forecasts for observations. We presented tests of a filament extraction method on gas and mock galaxies, as well as comparisons between detections in projected 2D and 3D positions. In the current paper, we take this one step further and consider finding filaments in observed redshift space. We focus on the specific challenge of finding filaments around clusters based on an observed three-dimensional distribution of mock galaxies.

\section{Motivation} 
\label{sec:problem}

In observations of bound structures like galaxy groups and clusters, observed positions along the line of sight get distorted and elongated due to peculiar velocities leading to inaccurate distance measurements. In addition, the amplitude of redshift space distortions differs depending on the galaxy type and redshift \citep{Coil2013}. Ultimately, this makes exploring the effects operating near and beyond the virial radii of galaxy clusters a challenging task. 

The phenomenon dubbed the “Fingers of God” \citep[FoG,][]{TullyFisher78} can be explained by considering that galaxies within a virialized structure at scales of $\sim1~$Mpc have large random motions relative to one another. Therefore, even though the cluster galaxies have similar distances to the observer, they have different redshifts. As a result, long fingers apparently extend from the cluster along the line of sight of the observer. For a massive cluster of $\sim$1400~km/s, this amounts to fingers stretching $\sim20\hMpc$ in each direction (Fig. \ref{fig:problem}).
Distortions arise also on scales larger than $\sim1~$Mpc, where the motions related to the infall of galaxies into collapsing clusters lead to an apparent contraction or flattening along the line of sight. This so-called Kaiser effect is therefore an opposing effect to the FoG and caused by many nearby galaxies all moving in the same direction, typically towards the centre of the cluster \citep{Kaiser1987}. 
Both distortions affect the cosmic web reconstruction significantly. 

We demonstrate the problem of extracting filaments in redshift space in Fig. \ref{fig:problem}, where we contrast clustering in real and redshift space. 
On the left, galaxies are plotted using true positions in three dimensions -- inaccessible to observers -- and on the right we show the same galaxies in redshift space with large FoG distortions pointing to the hypothetical observer, whose position we mark with a black symbol. The redshift distortions dramatically modify the geometry of the underlying field, so it is expected and not surprising that a 3D filament extraction in redshift space does not resemble the true filament network. We show this in Fig. \ref{fig:problem} where -- for the purpose of this demonstration -- the same filament extraction method is used on 3D galaxy positions on the left and distorted FoG simulations on the right\footnote{We followed the method described in \citetalias{Kuchner2020}, where we use halos with masses $M_{\rm{halo}} > 3\times 10^{10} \hMsun$ as tracers of filaments, extracted with \disperse\ using a persistence threshold $\sigma =6.5$ and a mass-weighting. See Sec.s \ref{sec:sample} and \ref{sec:disperse} for details.}.

While redshift space distortions are useful tools to reveal the underlying matter density and motions of galaxies, they complicate the extraction of filaments and hamper a measurement of the two-point correlation function in real space \citep{Coil2013}.
In our attempt to prepare for upcoming spectroscopic surveys of galaxy clusters and their outskirts such as the WEAVE Wide-Field Cluster Survey (WWFCS, Kuchner et al. in prep), we therefore need to investigate how peculiar velocities will impact the filament extraction. 

In the past, observers extracting the filaments of the cosmic web on large scales have either corrected for this effect statistically by compressing or truncating the FoG, before extracting filaments in 3D -- as was successfully demonstrated on very large scales \citep[e.g.,][using the SDSS and GAMA surveys]{Tegmark2004,Jones2010,Kraljic2017}, and on super-cluster scales \citep[][using the SDSS survey]{Santiago-Bautista2020} -- or reassigned filament segments with an excess of alignment \citep[][also using GAMA survey data]{Welker2019}. Alternatively, authors have resorted to extracting filaments based on a 2D projection of galaxies located in appropriate slices based on photometric redshifts, as was done by e.g., \citet{Laigle2017, Sarron2019}. Observers are therefore confronted with a choice: use the full 3D information including redshifts (and potentially correct for the FoG effect), or use a 2D projection. In \citetalias{Kuchner2020}, we have shown that the latter is a suitable option to finding filaments in cluster outskirts, if spectroscopic redshifts are provided to confine the cluster outskirt volume. By comparing filament extractions based on simulations of cluster volumes, we demonstrated that 2D filaments closely match their projected 3D counterparts. The goal of the current paper is to quantify the effect of redshift space distortions on filament finding and test whether a radial compression of cluster centres and groups in cluster outskirts presents an improvement on the 2D filament extraction discussed in \citetalias{Kuchner2020}.



\section{Simulations and Methods}
\subsection{Mock observations from \threehundred clusters}
\label{sec:sample}

As in \citetalias{Kuchner2020}, we use 324 simulations centered on massive galaxy clusters including their immediate surroundings out to a radius of $15\hMpc$ from \threehundred project\footnote{https://the300-project.org}\citep{Cui2018}. Briefly, the cluster simulations are zoom-in resimulations of the 324 most massive clusters at $z=0$ in the dark matter-only MDPL2 MultiDark simulation \citep{Klypin2016} that uses Planck cosmology ($\Omega_{\rm{M}} = 0.307, \Omega_{\rm{B}} = 0.048, \Omega_{\rm{\Lambda}} = 0.693, h = 0.678,$ \mbox{$\sigma_8 = 0.823$}, $n_s = 0.96$ \citep{Ade2016}) . To achieve this, all particles within a sphere with radius of $15\hMpc$ from the cluster centre at $z=0$ were traced back to their initial positions. These particles were split  into  dark matter and gas  particles with masses accordingly to the assumed cosmic baryon fraction. Outside this sphere of interest, low-resolution dark matter only particles with variable masses  mimic the effects of the large scale structure at larger distances. The final simulation suite consists of 129 snapshots at different time steps of each zoomed-in Lagrangian region, simulated using the  TREEPM+SPH code \gadgetx\ \citep{Beck2015, Rasia2015} with full physics galaxy formation modules. A more detailed description of the simulations is available in \citet{Cui2018}.

In \citetalias{Kuchner2020}, we introduced a sample of mock galaxies that was inspired by upcoming surveys of galaxy clusters and their outskirts. Most notably, mock galaxies are tailored to mimic observations from the Wide-Field Cluster Survey (WWFCS), but are equally applicable to the 4MOST cluster survey \citep{Finoguenov2019} and other planned surveys. These surveys are designed for detailed studies of the galaxy distribution in cluster infall regions, unveil pre-processing mechanisms considered to be responsible for observed environmental trends and research their constraints to cosmology.
With this in mind, we construct a mock galaxy sample using halos identified by the AHF halo finder \citep{Knollmann2009,Knebe2011}, which considers gas, stars and dark matter self-consistently. Halo properties like luminosity, stellar mass and peculiar velocity are based on the bound particles that account for a halo. We select halos in z=0 clusters that mimic the number counts and observable properties used to select galaxies for the WWFCS Survey: we aim for 4000 -- 6000 galaxies per cluster structure within 5\R\ with stellar masses roughly $M*>10^9\hMsun$. We refer to these halos as mock galaxies. 

This data set is suitable to answer the question we pose in section \ref{sec:problem}. The AHF catalogue provides simulated peculiar velocities [Vx,Vy,Vz] at the positions [x,y,z] of each halo. To simulate redshift distortions, we translate the peculiar velocities along the line of sight  -- arbitrarily chosen to be in the direction of the y-axis -- into distances and perturb each mock galaxy with this number (proportionate to its velocity relative to the centre). As a result we obtain galaxies positioned at [x,y$_{\rm{p}}$,z], where y$_{\rm{p}}$ stands for perturbed in y direction. We thus mimic Finger-of-God-like perturbations for all simulation volumes (Fig. \ref{fig:problem} shows one example) with the aim to investigate the quality of filament finding in 3D space, approximating observations. 


\begin{figure}
   \centering
  \includegraphics[width=\columnwidth]{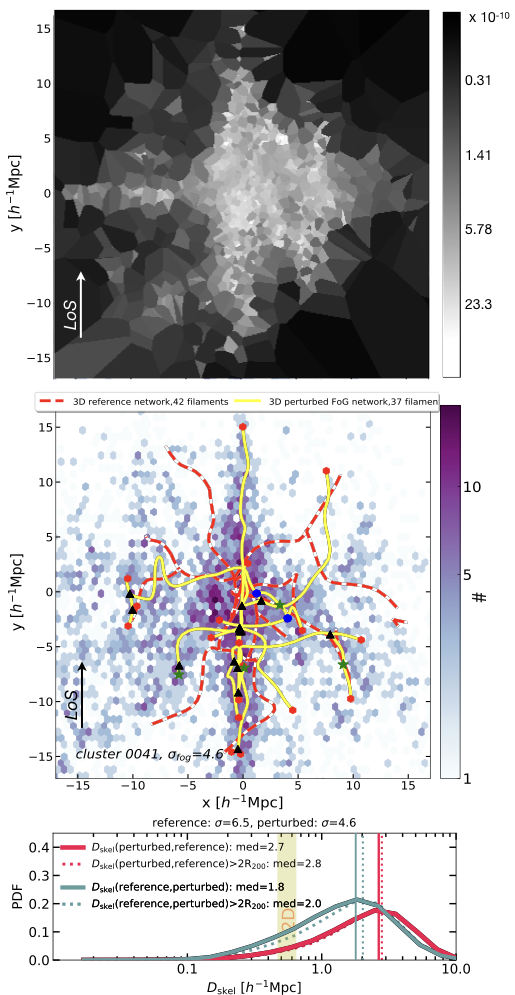}
  \caption{Top: Delaunay tessellation based on mock galaxies perturbed by the redshift space distortion used by \disperse\ to identify filaments, in a slice of thickness 75~kpc around the centre of one example cluster. Fingers of God structures along the LoS are clearly visible and picked up by the filament finding algorithm. This is shown in the yellow network in the middle panel. Blue dots mark nodes, green stars saddle points and black triangles bifurcations. As a comparison, the red dashed network is the result of running \disperse\ on the true 3D positions. The background shows the 2D histograms of the mock galaxies in redshift space. Bottom: Probability distribution of the distances between skeletons of filament networks from perturbed and true mock galaxies for the entire sample of 324 clusters. Pink curves show the distance measured from perturbed skeleton to the reference skeleton, blue curves show the distance from the reference skeleton to the perturbed skeleton. We test a range of $\sigma$-thresholds (see text) and show a result where both the numbers of filaments in each skeletons and their median distance converge. Dotted lines use segments outside $R_{200}$, dashed lines include them. Vertical lines show medians of the distances between filament extractions, values are printed in the legend.}
  \label{fig:fog_disperse}%
\end{figure}

\subsection{Filament finding with \disperse}
\label{sec:disperse}
In this work we use the topological filament extraction code \disperse\ \citep{Sousbie2011}. The code identifies topologically significant features in tesselated density fields that are calculated from an input of discrete positions of -- in the present case -- mock galaxies, either in 3D or 2D. The final network is constructed as a number of small segments that trace the ridges of the density field, referred to as \textit{skeleton}, as well as topologically robust critical points, i.e., saddle points and nodes. The input of a signal-to-noise criterium allows the user to recover a robust network with control over the scales at which filaments are found. This persistence measure (often expressed in terms of standard deviations $\sigma$ of a minimal signal-to-noise ratio) refers to the ratio of the value of two critical points in a topologically significant pair. Depending on the application, \disperse\ therefore offers oversight over whether faint tendrils should be included (with a tradeoff of increased noise) or if the analysis should focus on large scale, collapsed cosmic filaments. The scientific questions we ask in this body of work relate to galaxies in large filaments. In \citetalias{Kuchner2020} we compared mock galaxy networks to networks based on the underlying gas distribution and found a persistence threshold of $\sigma=6.5$ for the reference network to be appropriate. 

To find the appropriate persistence $\sigma$ in the present work, we iterate through a series of \disperse\ runs with a range of steadily increasing persistence thresholds from $\sigma =3$ to $\sigma=6.5$. When we assess the goodness of the network under review, we define the most successful setup as the network where (1) the number of filaments in the comparing network (i.e., the \textit{predicted} network) converges to the number of filaments of the reference network (i.e., the \textit{true} network), (2) the median distance between skeletons $D_{\rm{skel}}$ is small\footnote{$D_{\rm{skel}}$ is defined as the probability distribution function (PDF) of the distances between each segment of the predicted framework and its closest segment in the true framework \citep{Sousbie2009}} and (3) the distributions of both projections (i.e., $D_{\rm{predicted,true}}$ and $D_{\rm{true, predicted}}$) converge \citep{Malavasi2016}. However, in the present case where we use a fixed reference network, this last point is less important.\footnote{Tuning the goodness measure and finding the best parameters presents an obvious route into machine learning. While we do not explore this possibility here, \threehundred\ sample offers a suitable playground for machine learning algorithms. } Therefore, our measure of goodness relies on a combination of the number of filaments and the Dskel distributions, both in terms of reducing Dskel in general and minimising the distance between the means of the distributions. 
For further explanations of the extraction code, we refer the reader to \citet{Sousbie2011}, while a more detailed description of how we applied \disperse\ to extract filaments and compute $D_{\rm{skel}}$ can be found in \citetalias{Kuchner2020}.

\subsection{Filaments in redshift space}
\label{subsec:filaments_in_redshift_space}

The top panel of Fig. \ref{fig:fog_disperse} shows the \disperse-view of an example cluster with mass of $1.3 \times 10^{15} \hMsun$ in redshift space. The black-and-white image visualises the Delaunay tesselation in a slice of 75~kpc about the center of the cluster used by the algorithm to find filaments.  Like in Fig. \ref{fig:problem}, we see that the FoG structure dominates the centre. However, this system also shows the presence of a group in the same slice and thus a second FoG is picked up.

In the middle panel of the figure, we show the 2D histogram of the entire mock galaxy population in redshift space in the background and two networks in superposition: the yellow skeleton is the best filament network found for mock galaxies in redshift space. The red-dashed skeleton is our reference framework, based on the true positions of the same galaxies. We assess the extraction in the bottom panel, cycling through a range of persistence thresholds and considering both projections, as described in the previous section.
The user's choice of a persistence threshold greatly influences the nature of the network. Removing low persistence pairs (simplification) is the algorithm's main way to filter noise or remove “non-meaningful” structures. 
Following \citet{Sousbie2011}, \citet{Laigle2017} and others, and described in detail in \citetalias{Kuchner2020}, we compute $D_{\rm{skel}}$, the distances between the two skeletons we wish to compare, in all 324 cluster volumes and plot their differential distributions (PDF). The solid line is the PDF of distances of the sum of all skeletons (i.e., using all segments from 324 clusters), the dotted line is the result for segments outside $R_{200}$. The vertical lines indicate the corresponding medians, which range from 1.8 -- 2.8 $\hMpc$. Because the volume inside $R_{200}$ is small, medians that include segments inside $R < R_{200}$ are always lower. This is because inside $R_{200}$, segments lie close to each other. Comparing these numbers with other realizations of filament extractions offers a quantitative assessment of how close the method is to the ideal case of having knowledge of true positions. 

In the following sections, we will test whether we are able to recover the reference network after correcting for the FoG effect. As a measure of success, we will compare results of the corrected frameworks to the network we recovered from the 2D projections of galaxies (i.e., $D_{\rm{skel}} = 0.5$ and $0.65 \hMpc$, see \citetalias{Kuchner2020}) and marked as yellow bands in the plots. For reference, the distance of two random skeletons in our sample (i.e., comparing randomly selected pairs of cluster networks) leads to $D_{\rm{skel}} = 3.3$ and $3.8\hMpc$ for all segments and segments outside of $R_{200}$ respectively.


\section{Compression of virialized structures}
\label{sec:compression_steps}

\begin{figure}
   \centering
  \includegraphics[width=\columnwidth]{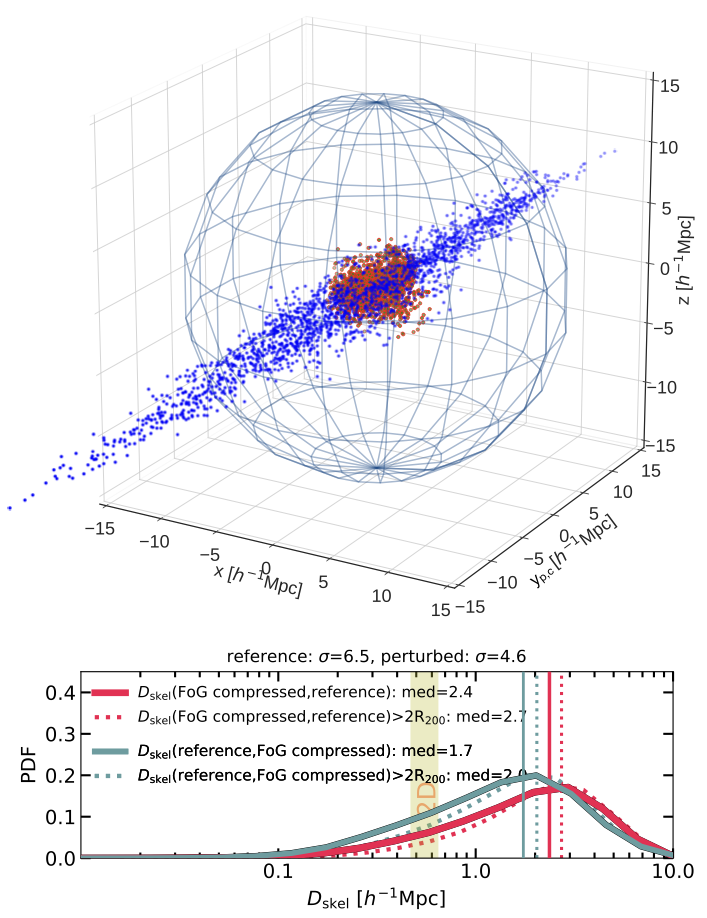} 
  \caption{In order to correct for the Fingers of God effect, we compress distorted mock galaxies of virialized structures. Top: First, we identify and normalize all perturbed galaxies inside 2\R\ of the main cluster halo (blue points), proxy for the cluster itself. The result is a compressed FoG (red points). We show cluster 1 of 324 as an example. The blue sphere indicates $15\hMpc$, roughly equivalent to 5$R_{200}$. Bottom: Assessment of \disperse\ filament extraction on corrected FoG data over the whole ensemble of 324 clusters. Probability distribution of distances between two skeletons in each cluster: one is extracted from true [x,y,z] positions of mass-weighted mock galaxies, the other from a perturbed [x,y$_{\rm{p}}$,z] distribution as in observed redshift space, but with the central 2$R_{200}$ compressed [x, y$_{\rm{c}}$,z] as explained in sec. \ref{sec:R200_compression}. Solid line: distance between all segments of the skeletons, dotted line: only segments outside $R_{200}$ are compared. Median distances are indicated in the legend. Distances are calculated from both projections (i.e., $D_{\rm{skel}}[\rm{predicted,true}]$ in pink and $D_{\rm{skel}}[\rm{true,predicted}]$ in blue).}
  \label{fig:step1}%
\end{figure}

The recent publications by \citet{Kraljic2017} and \citet{Santiago-Bautista2020} are based on observations in large volumes of up to several hundred $\rm{Mpc}^3$. Both studies extracted filaments after radially compressing dense regions -- groups in the case of \citet{Kraljic2017} and clusters in the case of \citet{Santiago-Bautista2020}.
However, solutions for traditional large-scale structure surveys that extend over several hundred Mpc may not be applicable in the case of finding filaments in the immediate vicinity of clusters where the field of view only extends a few virial radii of the cluster. In large surveys, $\sim 5\%$ of all filaments suffer from FoG effects of groups \citet{Welker2019}, whereas filaments in cluster outskirts are much more commonly affected (see Fig. \ref{fig:problem}).   
In addition, as \citet{Welker2019} pointed out, a correction based on the compression of groups can only be applied for rich groups ($\gtrsim$10 members). Moreover, it does not account for boundary effects that can cause spurious filaments along the line of sight. Another drawback of a FoG compression is that it also removes true filaments oriented along the line of sight. However, this can be accounted for in large volumes \citep[e.g.,][]{Jones2010}.

A step closer to the special case of cluster outskirts is the recent study by \citet{Santiago-Bautista2020}: their study focuses on a sample of 46 supercluster volumes in the local universe using SDSS-DR13 data where they selected superclusters with five or more clusters with box volumes between $\sim90$ and $3900 h^{-3} \rm{Mpc}^{-3}$. The authors also used a list of filament candidates -- identified as chains of at least three clusters -- as a prior to select the superclusters with the most promising filaments. They identified galaxy systems (large groups and clusters) in each of the 46 supercluster complexes before applying a virial approximation to correct the positions of the galaxies through scaling the comoving distances along a cylinder of radius $R_{\rm{aperture}}=R_{\rm{vir}}$ to the calculated virial radius.
The statistical compression of the FoG used in both examples \citep[i.e.,][]{Kraljic2017, Santiago-Bautista2020} therefore assumes virialization for group and cluster-sized nodes, thus imposing that the distributions in positions inside the groups are isotropic. This is not the case for cluster outskirts.

Following the successful method of correcting FoG distortions presented in \citet{Kraljic2017} and \citet{Santiago-Bautista2020}, we identify suitable areas in the simulated volumes that are most strongly affected by redshift distortions. 
In observations, one would have to find centres of clusters and groups -- \citet{Robotham2011} gives an example of how this can be determined observationally -- and estimate a suitable radius experimentally. While it is important to note that uncertainties are large, especially for lower mass systems \citep{Old2014}, this can be done e.g., through a virial approximation as is explained in \citet{Santiago-Bautista2020} and based on an algorithm presented in \citet{Biviano2006}. All galaxies projected inside the cylinder of this aperture radius and with velocities deviating significantly from the mean cluster velocity (to determine the length of the cylinder along the line of sight) can thus be defined as FoG galaxies.
The aim of the experiment we present in this paper is to test the "best case scenario" using simulations, where we benefit from the a priori knowledge of true positions, cluster properties such as $R_{200}$ and a limited volume. We are therefore able to determine members of virialized structures based on the true positions [x,y,z] of mock galaxies, i.e., an idealized case. We consider a filament extraction successful if they improve the 2D classification.

\begin{figure}
\centering
\subfloat[Identification of groups outside 2R$_{200}$]{%
  \includegraphics[width=0.35\textwidth]{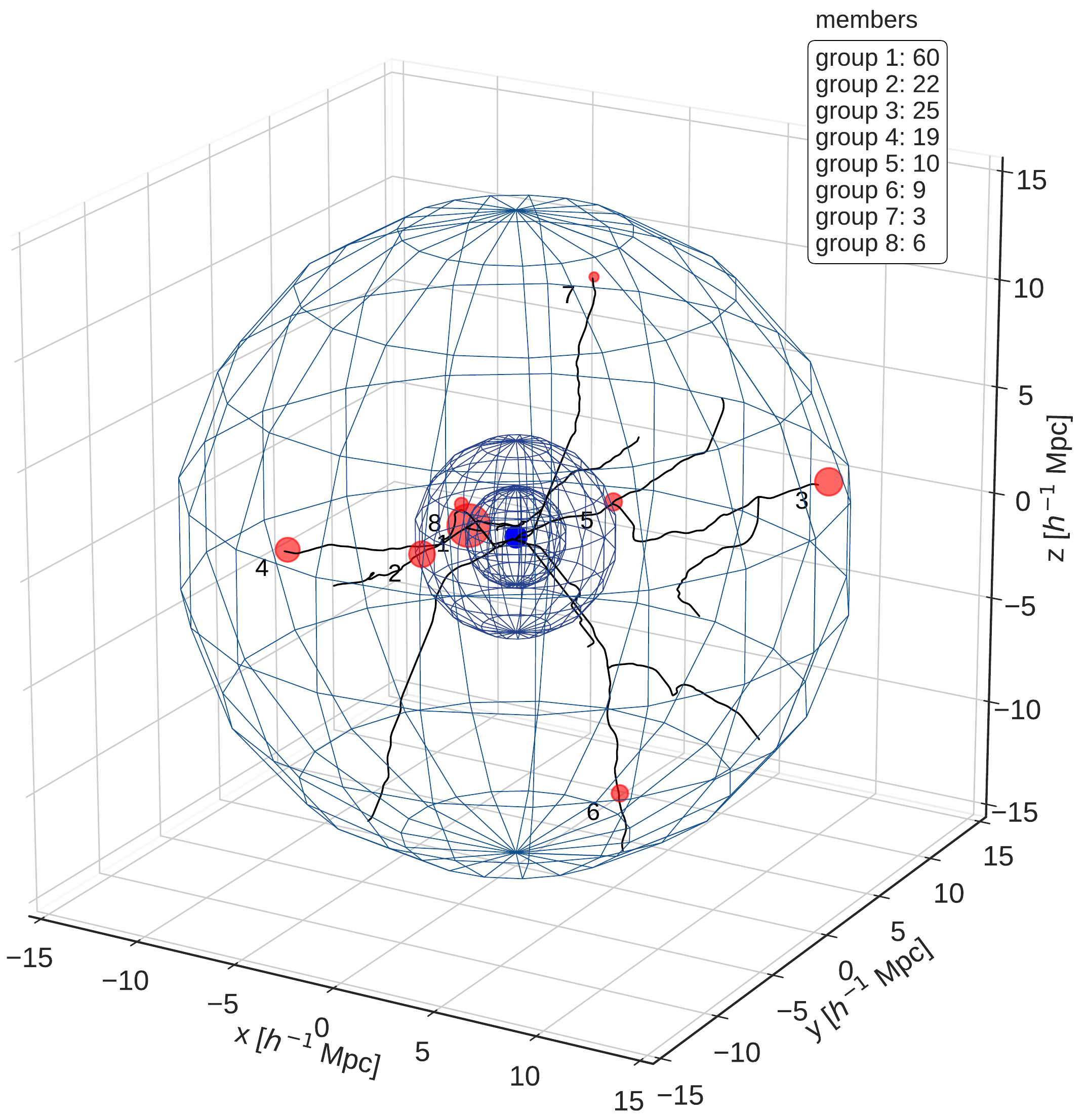}\label{fig:step2}%
  }\par        
\subfloat[Compression of group FoG]{%
  \includegraphics[width=0.35\textwidth]{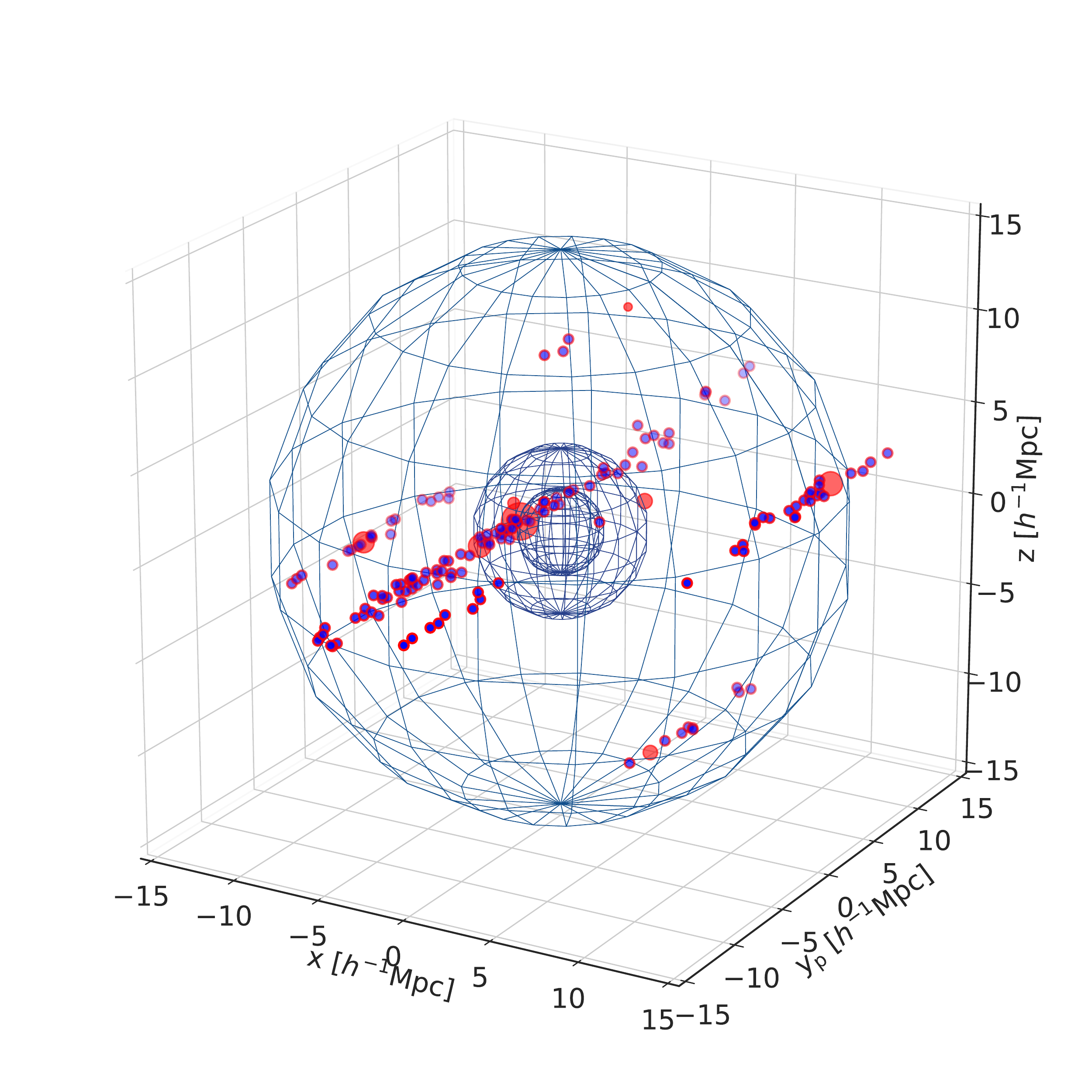}\label{fig:step3}%
  }
\caption{In a second step, we identify groups based on halos outside 2$R_{200}$ of the cluster with high velocity dispersions and find members within $R_{200}$ of that host halo (panel a). Then, we compress the FoG distortions from these groups (blue filled circles in panel b) in addition to cluster centres as shown in Fig. \ref{fig:step1}. The blue spheres indicate 1, 2 and 5 \R\ of cluster 1 of 324 as an example. Large red spheres indicate the position of groups, their sizes are scaled to the number of members which can also be read out in the insert of the top panel.}
\label{fig:steps_compression}
\end{figure}

\subsection{Compression of cluster centres}
\label{sec:R200_compression}

The densest regions in our cluster volumes, and therefore areas most affected by FoG distortions, are the cluster centres. Using  true positions [x,y,z], we define cluster centres as galaxies within 2\R\ of the main halo of the AHF catalogue, where $R_{200}$ is the radius of a sphere where the mean density is 200 times the critical density of the Universe. In \threehundred simulations, we use the main halo as a proxy for the cluster itself and its position as the centre of mass. 
In the simulated observations, all galaxies are redshift distorted [x,y$_{\rm{p}}$,z] as explained in Sec. \ref{sec:sample}. 

To test the impact of correcting for the FoG of the cluster centre, we compress the (perturbed) selected cluster centre galaxies radially. This is done by scaling the y$_{\rm{p}}$ positions of each FoG galaxy radially to inside $2R_{200}$. For each galaxy cluster, we thus normalize all FoG galaxies to within $2R_{200}$ of the cluster. In appendix \ref{sec:appendix_compress}, we assess this process and show that the compressed galaxies resemble the distribution of true central galaxy positions adequately well in the context of this exercise.
All galaxies contributing to the FoG within a radius $2R_{200}$ are thus re-distributed to [x, y$_{\rm{c}}$,z], where y$_{\rm{c}}$ stands for compressed, in a sphere of radius $2R_{200}$ about the centre. All other galaxies in the simulation box remain perturbed, as they would appear in observations. 
Fig. \ref{fig:step1} gives an example of the isolated FoG from all mock galaxies (i.e, galaxies obtained according to the survey’s selection function) within $2R_{200}$ (blue dots). The buildup of red points in the centre are the same mock galaxies, but now compressed to within a sphere with radius $2R_{200}$. Note that $2R_{200}$ does not always correct \textit{every} "FoG galaxy". We also tested whether other choices could improve the result and considered compressing galaxies within 1$R_{200}$, as was suggested by \citet{Santiago-Bautista2020}. We found that this was not large enough. Extending the radius even further, however, would go too far considering the whole field of view only extends out to 5$R_{200}$. Twice $R_{200}$ was thus chosen as a suitable compromise. 

To assess whether the compression of FoG galaxies improves the filament extraction, we compare locations of extracted \disperse\ skeletons as before (see Sec. \ref{subsec:filaments_in_redshift_space}).
This is shown in the bottom panel of Fig. \ref{fig:step1}, that quantifies the discrepancy/similarities between the reference mock galaxy-skeleton based on true positions [x,y,z] and skeletons retrieved after the FoG compression \textit{over the whole ensemble of clusters}. This means that we extract filaments from mock galaxies that have been perturbed according to redshift distortions [x,y$_{\rm{p}}$,z], but "FoG galaxies" are compressed to [x, y$_{\rm{c}}$,z] and compare the result to the reference network. Here, we show the result of our procedure to finding the best fit for the persistence threshold, in this case $\sigma=4.6$. 
The medians in Fig.~\ref{fig:step1} reveal that segments of the two networks are roughly $2 \hMpc$ apart, which means that this is not an adequate solution to the problem and does not lead to any significant improvement over the original FoG scenario shown in Fig.~\ref{fig:fog_disperse}. For comparison, in \citetalias{Kuchner2020}, we found good agreements between networks of different tracers when $D_{\rm{skel}}$ is between 0.5 and 0.7$\hMpc$.

\begin{figure}
	\includegraphics[width=0.95\columnwidth]{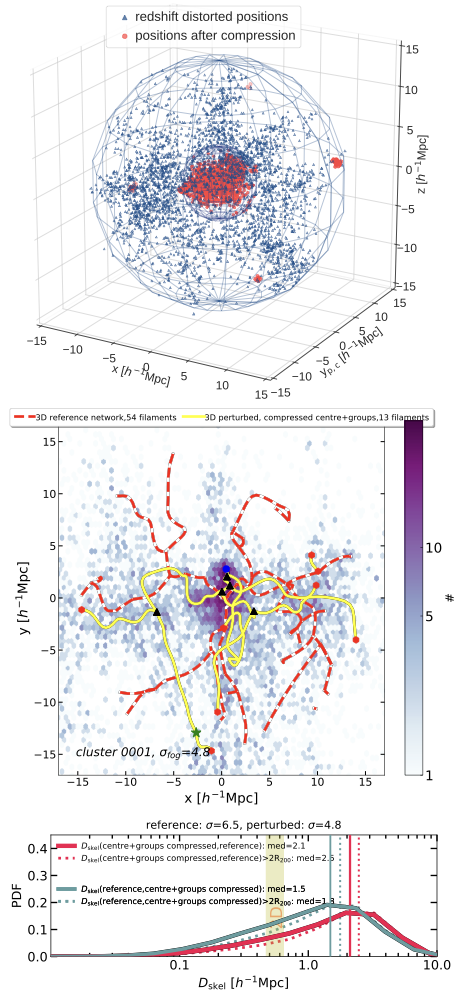} 
    \caption{Top: example cluster 0001 -- the most massive of \threehundred simulations -- showing compressed groups and compressed central $2R_{200}$ (red points) while the rest of the galaxies are perturbed in the y direction according to redshift distortions (blue points). The outer blue mesh shows a sphere of $15\hMpc$ radius, roughly comparable to 5$R_{200}$, centred on the main halo of the cluster. Middle: comparison of extracted filaments in this example cluster. One is extracted from true [x,y,z] positions of mock galaxies as before (our reference filament network in red-dashed lines), and the other is based on the  perturbed [x,y$_{\rm{p}}$,z] galaxy distribution as in observed redshift space but with the central 2$R_{200}$ as well as all groups compressed [x, y$_{c}$,z] (in yellow). Nodes were compressed before extraction. The background shows the 2D histogram of the galaxy distribution. Nodes, saddle points etc. are marked as described in Fig. \ref{fig:fog_disperse}. } Bottom: the PDF of the distances between segments of the skeletons over the whole ensemble of clusters shows only minor improvements.
    \label{fig:step2_combined}
\end{figure}

\subsection{Group identification and compression}
\label{sec:group_compression}
Similar to Fingers of God around cluster centres, we also find an artificial stretching of groups along the line of sight resulting in the creation of spurious filaments in the same direction.
As a next step to improving filament finding in redshift space, we therefore apply the same compression procedure not only for cluster centres, but also for galaxy groups. Fig. \ref{fig:steps_compression} shows the step of identifying groups (middle panel, \ref{fig:step2}) and what the FoG look like on group scales (lower panel, \ref{fig:step3}). 

We first identify group centres by looking for halos outside 2$R_{200}$ of the cluster with high (1D) velocity dispersions ($\sigma$ > 300~km/s). Each halo within $1R_{200}$ of this host halo is tagged as a group member. Note that this definition is different to identifying groups based on their halo/sub-halo status known in the simulations \citep[e.g.,][using the same data set]{Arthur2016a} and yields a smaller number of groups. 
Classifying groups based on sub-halo status relies on knowledge of whether halos lie within a common isodensity contour, with a subsequent removal of objects outside the host's R$_{200}$. With an eye to observations, groups may be more easily defined based on their velocity dispersion as demonstrated here, or other more commonly used methods, such as the Friends-of-Friends method. The exact choice of group identification method should be based on the individual science case and available data, and does not alter our conclusions. 

In \threehundred volumes, we find groups of various richness (indicated by the size of the red spheres in Fig. \ref{fig:steps_compression}). In ten regions of 324, a second cluster with \mbox{M$_{200} > 10^{15} \rm{h}^{-1} \hMsun$}, where M$_{200}$ refers to the mass enclosed within a sphere of radius R$_{200}$, is found in addition to the main cluster. In this exercise, we treat the second cluster in the same way as all groups. Therefore, some groups may host several hundred group members, others only a few. We compress each group's FoG (Fig. \ref{fig:step3}) as described in the previous section and run \disperse\ on the mock galaxies sample that is constituted of galaxies that are (1) compressed if they are within the central 2$R_{200}$, (2) compressed if they are within group regions, i.e., within $1R_{200}$ of high velocity dispersion halos and (3) perturbed elsewhere (top panel of Fig. \ref{fig:step2_combined}). The resulting PDF can be seen in the bottom panel of Fig. \ref{fig:step2_combined}. 

Even in the case where in addition to the FoG from cluster centres, those from groups -- prominent nodes in the network -- are compressed, the extraction does not improve significantly and therefore does not outperform the extraction based on projected 2D positions.
This is evident from the large distances between the skeletons (We report a median distance of 1.3$\hMpc$ and 2.5$\hMpc$ for all segments and 1.5$\hMpc$ and 2.1$\hMpc$ for segments outside 1\R, Fig. \ref{fig:step2_combined}). The two values refer to the two projections of calculating $D_{\rm{skel}}$. As before, the same care was taken with updating the \disperse\ setup iteratively. 

Both the large distance between networks in the quantitative assessment shown in Fig. \ref{fig:step2_combined} as well as the qualitative inspection of overlaying networks confirms that compressing virialized structures in the small volumes of cluster outskirts is not sufficient to robustly find the same filaments in 3D observed space as in 3D true positional space. 
In summary, minimising the node contributions by compressing dense regions in cluster volumes has not improved the filament extraction significantly. The average distance between skeletons in observed redshift space and skeletons extracted after the compression process is $D_{\rm{skel}} \sim 0.6 \hMpc$. In appendix \ref{sec:appendix_additional_tests}, we discuss two alternative approaches: connecting groups and rejecting galaxies in high-density regions. Neither test was able to considerably improve the filament finding, let alone perform better than filament finding in 2D.

\begin{figure}
	\includegraphics[width=\columnwidth]{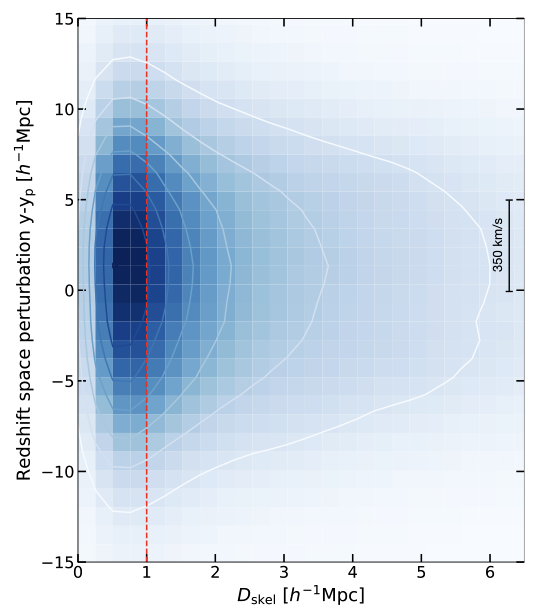}
    \caption{In observed redshift space, the positions of filament galaxies are apparently perturbed, similar to clusters or groups. The 2D density plot shows the distance from all mock galaxies outside 2\R\ and outside groups to their nearest filament segment in Mpc on the x-axis and perturbation along the line of sight on the y-axis. This visualization paints a clear picture of the magnitude of the perturbations of infalling cluster galaxies in redshift space.} The red dashed line indicates a characteristic thickness of a filament ($1 \hMpc$) as determined in \citetalias{Kuchner2020}.  The majority of mock galaxies are distorted by a few Mpc, but galaxies can be perturbed by up to $10 \hMpc$.
    \label{fig:perturbation}
\end{figure}

\section{Discussion}

In the previous sections we have seen that at the relatively small scales of cluster outskirts, finding filaments in redshift space may not be achievable using velocity information as a third dimension. The volumes are dominated by the main cluster, but even if we successfully correct for the kinematic artifacts from cluster centres and groups through a compression of FoGs, the smearing of the filaments themselves makes the filament finding with \disperse\ unreliable. The top panel in Fig. \ref{fig:step2_combined} gives a sense of this washing out of filaments, i.e., the perturbations of the filaments themselves. In redshift space, the apparent distribution of galaxies around filaments is much more diffuse than in true positions and appear elongated -- much like Fingers of God. Above the galaxy mass limit that we are using in this study ($\rm{M}_{\rm{halo}}>3\times10^{10}~\hMsun$ or $\sim\rm{M}_*>3\times 10^9~\hMsun$), these "filament FoG" dominate the cluster outskirt volume and are the reason why we struggle to find filaments in redshift space. 

\subsection{The perturbed filamentary network}

In Fig. \ref{fig:perturbation} we demonstrate the extent of these "filament FoGs" and thus quantitatively reveal the practical reasons for the difficulty of finding filaments in redshift space in the vicinity of clusters. The plot shows the distortion from true positions by peculiar velocities in Mpc/h as a function of distance to the filament network (D$_{\rm{skel}}$), also in Mpc/h. The plot is based on all mock galaxies outside 2\R\ and outside groups for all clusters combined. The kernel density estimation shows this distribution as a smooth indication of density in addition to 2D density contours. Note that there are noticeable cluster-to-cluster variations depending on the degree of substructure in each cluster volume. 
The red dashed line indicates a characteristic thickness of filaments, as determined by gas density profiles (see \citetalias{Kuchner2020}), implying that mock galaxies with a maximal orthogonal distance of 1$\hMpc$ to the skeleton are "inside" filaments. This is of course a simplified view, since filaments are diffuse constructions of cooled filament gas concentrated around the filament spine without a clear edge. However, it helps to associate galaxies to filaments and illustrates the apparent spread of filament galaxies. The 2D density contours reveal that the apparent positions of most mock galaxies inside filaments are typically perturbed by a few Mpc, with some up to $10 \hMpc$ ([y - y$_{\rm{p}}$]). Given we only probe a volume of radius $15 \hMpc$ around the cluster, this explains why our attempts to extract filaments in 3D using [x,y$_{\rm{p}}$,z] have failed and assessments show large discrepancies of around $2 \hMpc$ in comparison to filaments based on [x,y,z] positions (Fig. \ref{fig:step1} and \ref{fig:step2_combined}). 
As a consequence, \disperse\ is struggling to find filaments. 

Large scale deviations from real positions correlate with orientation to the line-of-sight (Fig. \ref{fig:problem}). We therefore investigated whether a specific orientation of filaments with respect to the line-of-sight might be responsible for the large perturbations. We extracted two sets of filaments, those parallel and those perpendicular to the line-of-sight within 20 degrees and allowing for curvature of filaments. We did this in the reference and the perturbed networks and compared each set. While, on average, there are comparatively more parallel filaments in the perturbed networks due to the stretched FoGs, sets of perturbed filaments in either orientation are uncorrelated to the corresponding reference filaments. This means that no specific orientation can be made responsible.

\subsection{The velocity field of filament galaxies}
\label{sec:velocity_field}

In order to understand the underlying reasons for this complication, it can be helpful to look at how matter flows between different morphological components of the cosmic web.
The vast observational Cosmic Flows program \citep[e.g.,][]{Courtois2011,Courtois2013, Tully2014} of peculiar velocities in our local Universe (up to \mbox{30,000 km/s}) has shown how mass flows along and towards structures of the cosmic web. Their detailed maps of observed (and reconstructed three-dimensional) motions of galaxies offer a valuable input to the translation from redshift space to physical space. It reveals that the local dynamics in both low- and high-density regions greatly impact the inferred density distribution that constitutes the input to a filament extraction from observations. 
The velocity information from Cosmic Flows offers clear signatures of flows that converge on major filaments and then progress toward peaks of the galaxy distribution\footnote{See also \citet{Ma2014} for a general overview of the cosmic velocity field and \citet{Trowland2012} for a discussion on the manifestation of the systemic bulk flows of filaments from numerical simulations.}. 
While clusters are undoubtedly the greatest basins of attraction for galaxies, matter thus also moves towards filaments.

\begin{figure}
	\includegraphics[width=\columnwidth]{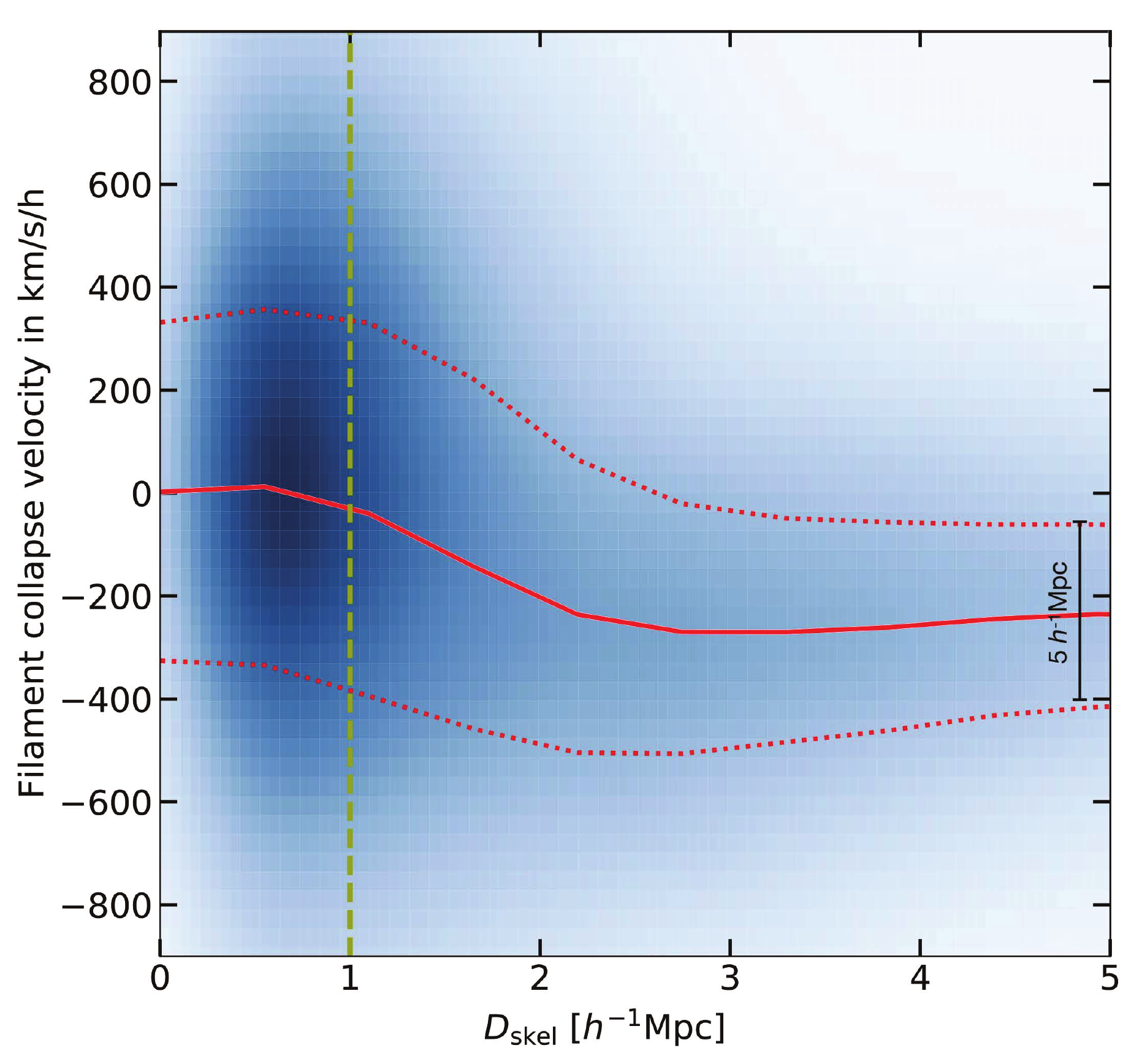} 
    \caption{Collapse velocity of mock galaxies towards filaments. This plot shows a typical "infall velocity" of $\sim$200 km/s of all mock galaxies outside 2\R\ and inside $10 \hMpc$ of the cluster, plotted as a function of distance to the filament spine (for filaments determined on [x,y,z] positions). The yellow dashed line indicates a characteristic thickness of filaments ($1\hMpc$) as determined in \citetalias{Kuchner2020} and red lines mark the median, lower and upper quartiles of the radial velocity component.}
    \label{fig:infall_vel}
\end{figure}

To further our understanding of the problem explored in this paper, and thus explain the perturbations shown in Fig. \ref{fig:perturbation}, we now focus on the small scale flow patterns initiated by the collapse of filaments\footnote{In this work, we describe streaming motions as motions in the rest-frame of the cluster, defined by the mean of all mock galaxies within \R. }. 
We therefore investigate how galaxies are flowing perpendicular to the reference filaments in 3D and separated from their movement towards the clusters. This movement is independent of the orientation of filaments. To achieve this, we first need to take out the bulk infall towards the centre of the cluster (i.e., the average motion of galaxies towards the cluster, assumed to be radial). This was done by averaging the velocity component toward the cluster in radial bins and correcting the velocity of each mock galaxy accordingly. 
Fig. \ref{fig:infall_vel} shows the resulting isolated radial velocity component towards filaments (i.e., orthogonal to each segment of the skeleton) of all mock galaxies in 324 simulations towards the filaments (extracted from true galaxy positions [x,y,z]) as a function of distance to the skeleton in Mpc/h. Negative velocities imply a movement towards the skeleton and the typical "thickness" of a filament is indicated by the yellow dashed vertical line (see discussion above). 
Very close to the centre of the filament, large random motions averaging at 0~km/s (comparable to a filament velocity dispersion of \mbox{$\sim$300 km/s}) make it impossible to isolate any collapse velocity. However, away from the spine of the filament, we detect a signal of negative velocities, i.e., velocities that indicate that galaxies are statistically falling toward the filaments, which is in addition to the movement towards the cluster centre.
Red lines mark the median (solid line) and lower and upper quartiles (dotted lines) of this collapse velocity, which we may envision as a "flow towards filaments". Following this median line from higher distances to lower distances (i.e., towards filaments), we see that galaxies "stream" towards filaments with an average collapse velocity of 200 km/s. 


Fig. \ref{fig:infall_vel} shows results of galaxy flows in highly complex environments. Firstly, we focus on z=0, where filaments could be relatively more disturbed than at earlier times. Cosmological hydrodynamical simulations show that at higher redshifts, satellite galaxies and gas fall nearly radially along well-defined cold filamentary streams to the centre of massive halos on extremely short time-scales \citep{Dubois2012}. Secondly, we focus on clusters and cluster outskirts, where filaments are exposed to highly mixed and turbulent environments, where the gas undergoes significant shocks \citep{Power2019} and filaments become relevant locations of collapse. Velocity flows in these exceptional regions in the Universe deserve a thorough discussion. In \citet{Rost2020a} we do exactly that and investigate velocity flows of gas and dark matter around clusters in much greater detail, using the same simulations. 
\section{Conclusions: selected usage of spectroscopic information}
\label{Conclusion}

Observations are dotted with obstacles to overcome. These include magnitude limits, volume limits, biases of multiple sorts, incompleteness, sparse sampling and the redshift space distortions we focused on in this paper. 
Studies that benefit from the large areas they cover  -- often with a tradeoff of higher redshifts -- like SDSS \citep{Yan2013,Martinez2015, Poudel2017}, GAMA \citep{Alpaslan2016, Kraljic2018} and VIPERS \citep{Malavasi2017} reconstruct the web in 3D by connecting high density nodes, usually galaxy groups, over large distances. On supercluster scales, mapping filaments can be achieved through identifying bridges between cluster pairs \citep{Cybulski2014} and elongated chain-like structures between dense galaxy systems \citep{Santiago-Bautista2020}.

Our study is narrowing down on environments where a filament reconstruction in 3D has not been successfully described before. 
This is likely due to the combination of the relatively small volume/region on the sky and the complex accumulation of largely unvirialized systems of galaxies that dynamically interact and flow between the features of the cosmic web. 
Accounting for the perturbations caused by the peculiar velocities and transforming them to real space has proven to be a challenge in this narrow regime: velocities can only be measured in the radial direction, distances have large uncertainties, and deviations from cosmic expansion are very uncertain for individual objects \citep{Courtois2013}. The method of defining the web kinematically by directly mapping galaxy peculiar velocity flows is so far restricted to the very nearby Universe \citep{Tully2014, Dupuy2019}.

How, then, can we trace filaments around clusters in observations if the valuable information of thousands of spectra of galaxies is not offering easy-to-obtain positions (distances) in the line of sight direction? 
Surveys that will attempt to map filaments in cluster outskirts may have to fall back to 2D reconstructions.
In \citetalias{Kuchner2020}, we discussed a comparison between filaments obtained based on 3D positions and 2D projections (2D positions on the sky) of mock galaxies. We saw that, to a large extent, the two methods trace the same cosmic structures with median D$_{\rm{skel}}$=0.51 $\hMpc$ for all segments and D$_{\rm{skel}}$=0.61 $\hMpc$ for segments outside of \R. This is far superior to the values that we obtained through the steps we undertook in the present paper. Using 2D projections therefore proved to be a good alternative in the presented case. 

In addition, \citetalias{Kuchner2020} provided numbers to evaluate the impact of projections on recovery rates for galaxies associated to filaments found in 2D compared to 3D (Sec. 3.5.2). We defined galaxies in filaments as galaxies with orthogonal distances smaller than 0.7$\hMpc$ or 1$\hMpc$. While \textit{apparently} there are more galaxies close to filaments due to the projection onto a plane, the contamination rate was relatively moderate: we found that 67\% (75\%) of all mock galaxies in filaments of thickness 0.7$\hMpc$ (1$\hMpc$) are still correctly identified in 2D. To put this into perspective, this true positive rate was still 5 times higher than if we randomly selected galaxies, where only 14\% are located in filaments.

In order to reach this quality and unambiguously study the projected filaments in cluster outskirts, narrow volumes around clusters need to be identified. All our tests were performed in a controlled volume of a sphere with 15$\hMpc$ radius around the cluster. Even high precision photometric redshifts will struggle to reach this level of accuracy. 
Only spectroscopic redshifts with higher accuracy and precision along the line of sight allow us to isolate the environment to the area of interest.
This will be possible for targeted cluster outskirt campaigns with instruments such as WEAVE and 4MOST.
Using WWFCS as an example, we expect radial velocities from WEAVE with uncertainties smaller than 25 km/s, and because peculiar motions induce distance errors of the order of many Mpc, spectroscopic redshifts will ensure that we only probe galaxies in the volume of 5\R\ radius around the cluster (comparable to $15\hMpc$). The large number of optical fibers and an optimised targeting strategy will lead to a high density of galaxies with spectroscopic redshifts corresponding to this volume. For WWFCS, we calculated that we will reach between 4000 and 6000 spectroscopically confirmed cluster structure members for each of the 16 clusters with at most $\sim$3\% incompleteness. Similar numbers can be expected for other cluster outskirt surveys.
This is the basis for successfully characterizing filaments connected to galaxy clusters in observations. Spectroscopic information will therefore still be vital for the selection of galaxies, however they will not be used directly for the input to finding filaments in the small region of cluster outskirts. For that, we will resort to a 2D extraction.

\section*{Acknowledgements}

We thank the referee for providing useful feedback to this study. This work has been made possible by \threehundred collaboration\footnote{https://www.the300-project.org}, which benefits from financial support of the European Union’s Horizon 2020 Research and Innovation programme under the Marie Sk\l{}odowskaw-Curie grant agreement number 734374, i.e. the LACEGAL project. \threehundred simulations used in this paper have been performed in the MareNostrum Supercomputer at the Barcelona Supercomputing Center, thanks to CPU time granted by the Red Espa\~{n}ola de Supercomputaci\'{o}n.
UK acknowledges support from the Science and Technology Facilities Council through grant number RA27PN. AK is supported by MICIU/FEDER under research grant PGC2018-094975-C21. He further acknowledges support from the Spanish Red Consolider MultiDark FPA2017-90566-REDC and thanks Ride for nowhere. GY acknowleges financial suport from  MICIU/FEDER through research grant number PGC2018-094975-C21.
ER acknowledges funding through the  agreement ASI-INAF n.2017-14-H.0
\newline
The authors contributed to this paper in the following ways: UK, AAS, MEG, AR and FRP formed the core team. UK ran \disperse, analysed the data, produced the plots and wrote the paper with ongoing input from the core team and valuable comments for improvement from collaborating co-authors. GY supplied the simulation data; AK the halo catalogues. 

\section*{Data Availability}
Data available on request to \threehundred collaboration.



\bibliographystyle{mnras}
\bibliography{references.bib} 



\appendix
\section{Assessment of the FoG compression algorithm}
\label{sec:appendix_compress}

\begin{figure}
	\includegraphics[width=\columnwidth]{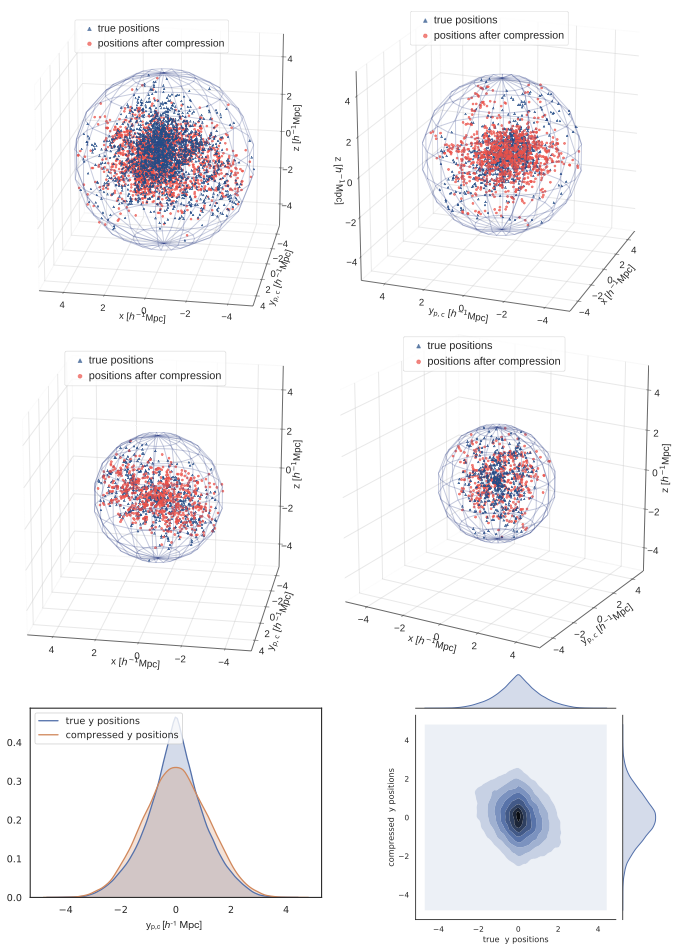}
    \caption{Assessment of the FoG algorithm: we compare true 3D positions of galaxies within 2$R_{200}$ (blue triangles) with the same galaxies perturbed in redshift space (FoG) and then compressed (red dots). The first four figures show example clusters showcasing the range of masses and morphologies covered by \threehundred\ in different angles. The bottom row shows two plots in which we contrast the distributions of true and compressed y-positions,our selected "line of sight" direction. While the distribution of compressed galaxies is slightly wider than the compact centres in true positions, the compression algorithm resembles the overall distribution sufficiently well for our purpose.}
    \label{fig:appendix_compress}
\end{figure}

In this appendix, we investigate whether the compression algorithm adequately reproduces the real positions of galaxies in dense environments. Fig. \ref{fig:appendix_compress} demonstrates that radially scaling the distances along the FoG to the virial radius represents the true positions in the cluster centre very well. We follow the approach by \citet{Santiago-Bautista2020}, but with the advantage that we know centre positions and size of $R_{200}$. This simple process is very similar to what was also used in \citet{Kraljic2017}. In Fig. \ref{fig:appendix_compress}, we show a comparison of true 3D galaxy positions (blue triangles) and positions of FoG galaxies after the compression algorithm (red points). In the first two rows, we print different rotations of four randomly selected cluster centres ($R<2R_{200}$) to showcase a variety of masses. Each plot depicts a different cluster and the blue sphere encompasses 2$R_{200}$ of that cluster. Overall, the distribution of red points is comparable to the distribution of blue triangles. The KDE plots in the lower row show the distribution of true y-positions and y-positions after compression in 1D and 2D over all 324 cluster centres combined. We chose y to be the direction of LoS. The compression leads to a slightly wider result, revealing that the algorithm is not able to fully reproduce the compactness of cluster centres. However, this is not our goal. We remind the reader that our aim is not to find filaments in this region. We aim to find reliable filaments in the outskirts, i.e., outside of this region. The goal is to create a node in the centre that can be identified by \disperse\ as a maximum. Note that close to the centre, filaments will overlap and we discussed this volume effect in \citetalias{Kuchner2020}. Therefore, we cannot comment on filament positions very close to the central node without correcting for the volume first. 

\section{Additional tests}
\label{sec:appendix_additional_tests}

\subsection{Connecting groups}
Large groups with many members are prominent nodes in the filament network around clusters. In our attempt to find possible ways to identifying filaments using 3D data in observations, we tested whether a filament network could be achieved by simply connecting the compressed groups and cluster centre (as explained in Sec. \ref{sec:compression_steps}), thus completely omitting any galaxies \textit{not located} in virialized structures for the \disperse\ extraction. While this may work on very large scales, we found that it does not work on the limited volumes we are investigating. This was true also when we expanded the group search far beyond the 5$R_{200}$ volume we study here. 
This approach simply excludes too many mock galaxies to resemble meaningful filament networks.

\subsection{Finding filaments without nodes}
\begin{figure}
	\includegraphics[width=\columnwidth]{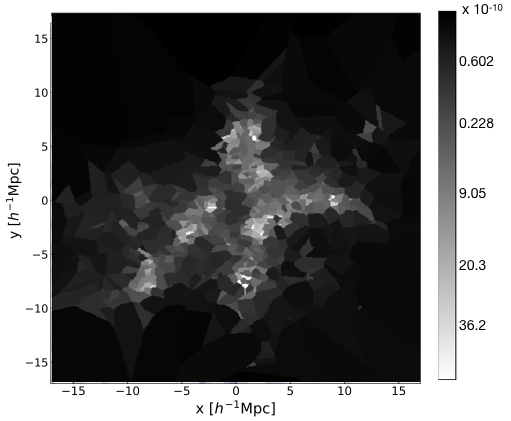}
    \caption{Removing galaxies in high density regions (within $2R_{200}$ of the cluster centre and groups) leaves an underdense/empty region. The figure shows the Delaunay tesselation which is the base for finding filaments wiht \disperse.}
    \label{fig:appendix_hole}
\end{figure}

In an additional approach we tested filament extractions after rejecting galaxies in high-density regions, i.e., galaxies located within groups and the cluster centre are removed prior to \disperse\ runs. The aim of this exercise is to omit all FoG "structures" of virialized regions, leaving the perturbed galaxies in the infall regions. As a result, \disperse\ interpreted the denser "shells" around these empty regions as dense ridges akin to filaments. The skeletons tend to follow ring-like features around empty central regions. Fig. \ref{fig:appendix_hole} helps to understand the issue. We calculate $D_{\rm{skel}} = 1.9 - 2.8~\hMpc$ in comparison to the reference skeleton. Therefore, at the small scales and detailed level that we seek to find filaments, this approach is no adequate solution. Filling this hole with a centralized concentration of mass to this (to find a maximum/node) is essentially the same as the approach of compressing FoGs into a ball in the centre. 

\bsp	
\label{lastpage}
\end{document}